\documentclass[fleqn,twoside]{article}
\usepackage[headings]{espcrc2}

\usepackage{epsfig}

\newcommand{\bc}{B_c}
\newcommand{\jpsi}{J/\psi}
\newcommand{\cdf}{{\tt CDF }}
\newcommand{\br}[1]{{\rm BR}(#1)}
\newcommand{\efig}[3]{
  \begin{figure}[htb]
    \begin{center}
      \epsfig{file=#3.eps,height=#1cm}
    \end{center}
    \caption{#2}
    \label{fig:#3}
  \end{figure}
}

\title{Studies of the $\bc$ meson at \cdf}

\author{Mario Spezziga
  \address[TTU]{Texas Tech University, Dept. of Physics, Box 41015,
    Lubbock, TX 79409-1051, USA}, for the \cdf collaboration
}
\runtitle{$\bc$ meson at \cdf}
\runauthor{\cdf collaboration}

\begin{document}

%
%
%

\begin{abstract}
We present the latest measurements of the $\bc$ meson properties using
360 $pb^{-1}$ of data collected by the \cdf detector. The results
include the $\bc$ mass and the ratio of branching fraction
$\bc\rightarrow \jpsi\: l$ with respect to $B\rightarrow \jpsi K$.

\end{abstract}

\maketitle

\section{INTRODUCTION}

 Within the standard model of elementary particles, five of the six
different kinds of quarks combine in quark-antiquark pairs to make
mesons.  The $B_c^{\pm}$ meson is the combination of the two heaviest
of these quarks, and is made of a bottom-charm antiquark-quark pair.
The CDF collaboration made the first observation of the $\bc$ meson in
the semileptonic decay channels $\bc \rightarrow \jpsi l \nu_l X$, in
a sample of 110 pb$^{-1}$ of data at $\sqrt{s} = 1.8$ TeV in Run I at
the Tevatron \cite{CDFRunI_bc}. With a signal of 20.4$^{+6.2}_{-5.5}$
events, the $\bc$ mass was measured to be 6.40 $\pm$ 0.39(stat) $\pm$
0.13(syst) GeV/$c^2$.  Recently, the D0 Collaboration reported a
preliminary observation of a $\bc$ signal in the decay channel $\bc
\rightarrow \jpsi \mu \nu_{\mu} X$ in a sample of 210 pb$^{-1}$ of Run
II data \cite{D0RunII_bc}. Up to now it has not been observed in any
fully reconstructed decay mode.  Consequently, its mass $M(B_c)$ has
not been measured with good precision.

In $p\bar{p}$ collision, the parton subprocess of gluon-gluon fusion
$gg(q\bar{q})\rightarrow(\bar{b}c)+b\bar{c}$ is the dominant process
for the $\bc$ production. In the leading approximation of QCD
perturbation theory, the calculation of subprocess for $\bc$ cross
section involves diagrams in the fourth order over the $\alpha_s$
coupling constant while the $gg(q\bar{q})\rightarrow b\bar{b}$ process
for the $B^+$ production involves $2^{nd}$ order over $\alpha_s$. The
production cross section of $\bc$ at Tevatron is thus predicted to be
rather low, in the order of $10^{-3}$ of that for $B^+$. Latest
calculations give a number of the order of 7.4 $nb$. The main channel
of $\bc$ meson decay occurs through the weak decays of the $b$-quark
or the $c$-quark, which are respectively about 25\% and 65\% of the
total decay width \cite{SaleevVasin,Kiselev_03}. The rest is given by
the annihilation decays $\bar{b}c\rightarrow l^{+}\nu, c\bar{s},
u\bar{s}$.

The CDF II detector consists of a magnetic spectrometer surrounded by
calorimeters and muon chambers and is described in detail
elsewhere~\cite{cdfIIdet}. The components relevant to this analysis
are briefly described here.  The tracking system is in a 1.4 T axial
magnetic field and consists of a silicon microstrip detector (L00,
SVX, ISL) \cite{cdfSVXII,cdfISL,cdfL00} surrounded by an open-cell
wire drift chamber (COT) \cite{cdfCOT}.  The muon detectors used for
this analysis are the central muon drift chambers (CMU), covering the
pseudorapidity range $|\eta|<0.6$ \cite{Ascoli,Dorigo}, and the
extension muon drift chambers (CMX), covering $0.6<|\eta|<1.0$.
Soft electrons are identified as tracks pointing to a cluster in the
Central Electromagnetic Calorimeter (CEM), with additional information
provided by the Central Shower Maximum Detector (CES), a proportional
wire chamber with 2-dimensional position measurement capability,
placed roughly at a depth of 6 radiation lengths in the CEM, near
shower maximum, and by the Central Pre Radiator (CPR) a plane of multi
wire proportional chambers situated in the gap between the solenoid
coil and the CEM, providing measurement in the x coordinate only.

These measurements use events containing muon pairs with $|\eta| <$
1.0, recorded with a three-level trigger, requiring pairs of muons
with opposite charge, $p_T>1.5 {\rm GeV}/c$ and invariant mass between
2700 and 4000 MeV/$c^2$. This will form the main ``$\jpsi$'' sample
for this analysis. For particle identification studies, samples
collected with a 2-track trigger ($D^0\rightarrow K\pi,
\Lambda\rightarrow p\pi$) and a single electron trigger
($\gamma\rightarrow e^+e^-$) are used.

\section{ANALYSIS}

The analyses presented here are divided into semileptonic decays
($\bc\rightarrow \jpsi \;l$ with $l=\mu$ or $e$) and fully
reconstructed decay $\bc\rightarrow \jpsi\pi$). The former has the
advantage of a larger branching fraction ($O(2\%)$) and statistics,
but it reduces essentially to a counting experiment with the necessity
of a careful determination of the background, while the latter
provides a precise determination of the mass, but a lower rate.

\subsection{Semileptonic channels}

The kinematical region for the semileptonic decay is the $M(\mu\mu l)$
range between the mass of the $\jpsi$ mass and the $\bc$
mass. Assuming $M(\bc)=6.4$ GeV we see that most of the signal events
lie between 4 and 6 GeV, which is defined as the search window. The
search consists in counting the candidates in this window and compare
them with the expected background events. The sources of background
are the following:
\begin{itemize}
\item Fake $\mu/e$ from $K/\pi/p$
\item Conversion electrons (only for the $\jpsi e$ channel)
\item $b\bar{b}$ events, with $b\rightarrow\jpsi +X$ and
  $\bar{b}\rightarrow e/\mu +X$
\item Fake $\jpsi$, i.e. unrelated muon pairs which appear to have a
  common vertex and invariant mass close to the $\jpsi$.
\end{itemize}
The fake $e\mu$ rate is estimated from the $D^*\rightarrow\pi^+
D^0\rightarrow\pi^+K^-\pi^+$ and $\Lambda\rightarrow p\pi$ samples,
where the $\pi$, $p$ and $K$ particles can be identified with
certainty and events counted by fitting the $D^0$ and $\Lambda$
invariant mass peaks. The probability for a $\pi$, $p$ and $K$ to be
identified as a lepton will be the ratio between the number of events
that passed all of the lepton identification cuts and the total number
of events in the sample. Once this rate has been measured, it is
multiplied by the probability for the third track in the vertex to be
a $\pi$, $p$ or $K$. This number is extracted from a fit of the
distribution of the Time Of Flight and $dE/dx$ data and from Monte
Carlo simulation.

The background from conversion electrons, i.e. those
$\gamma\rightarrow e^+e^-$ events where only one electron is
identified and associated to the muon pair, is estimated by evaluating
from a full simulation of the detector the efficiency for identifying
conversions. This efficiency is dependent on the momentum of
electrons. It is then normalized to the data sample.

Pair of $b$-quarks are produced by three processes: flavour creation,
gluon splitting and flavour excitation. These processes are generated
with the Pythia program \cite{Pythia} and passed through the full \cdf
simulation to evaluate the background from $b\bar{b}$ events, then the
number of such events is normalized to data, using the
$B\rightarrow\jpsi K$ sample. The relative importance of the three
processes in Pythia is checked with data, by fitting the angle
separation between the $\jpsi$ and the muon.

Finally, the amount of fake $\jpsi$ events is estimated from the
sidebands of the invariant mass distribution of the $\jpsi$.
\begin{table}[htb]
\caption{Summary of backgrounds in the semileptonic analyses}
\label{table:backsum}
\newcommand{\cc}[1]{\multicolumn{1}{c}{#1}}
\begin{tabular}{@{}lll}
\hline
&\cc{$\bc\rightarrow\jpsi\;\mu$}&\cc{$\bc\rightarrow\jpsi\;e$}\\
\hline
Fake $e\mu$           & $16.3\pm 2.9$        & $15.43\pm 0.31$   \\
$b\bar{b}$            & $12.7\pm 1.7\pm 5.7$ & $33.63\pm 2.20$   \\
Conversions           &      \cc{NA}         & $14.54\pm 4.38$   \\
Fake $\jpsi$          & $19.0\pm 3.0$        & \cc{negl.}        \\
Fake $\jpsi$ \& $\mu$ & $-2.0\pm 0.5$        & \cc{negl.}        \\
\hline
Total backg.          & $46.0\pm 7.3$        & $63.6\pm 4.9$     \\
Signal events         & $60 \pm 13$          & $115\pm 16\pm 14$ \\
Significance          & $5.2\sigma$          & $5.9\sigma$       \\
\hline
\end{tabular}
\end{table}
Table~\ref{table:backsum} shows the summary of all backgrounds with
their contribution, the number of signal events, and the significance,
defined as the probability that the background may fluctuate as much
as to reproduce the observed signal.

Figures \ref{fig:BcYield2} and \ref{fig:MuMuEMass} show the invariant mass
distributions of signal and background events for, respectively, the
$\bc\rightarrow\jpsi\; \mu$ and $\bc\rightarrow\jpsi\; e$ channels.

\efig{5.5}{Signal and background for the $\bc\rightarrow\jpsi\; \mu$
  channel}{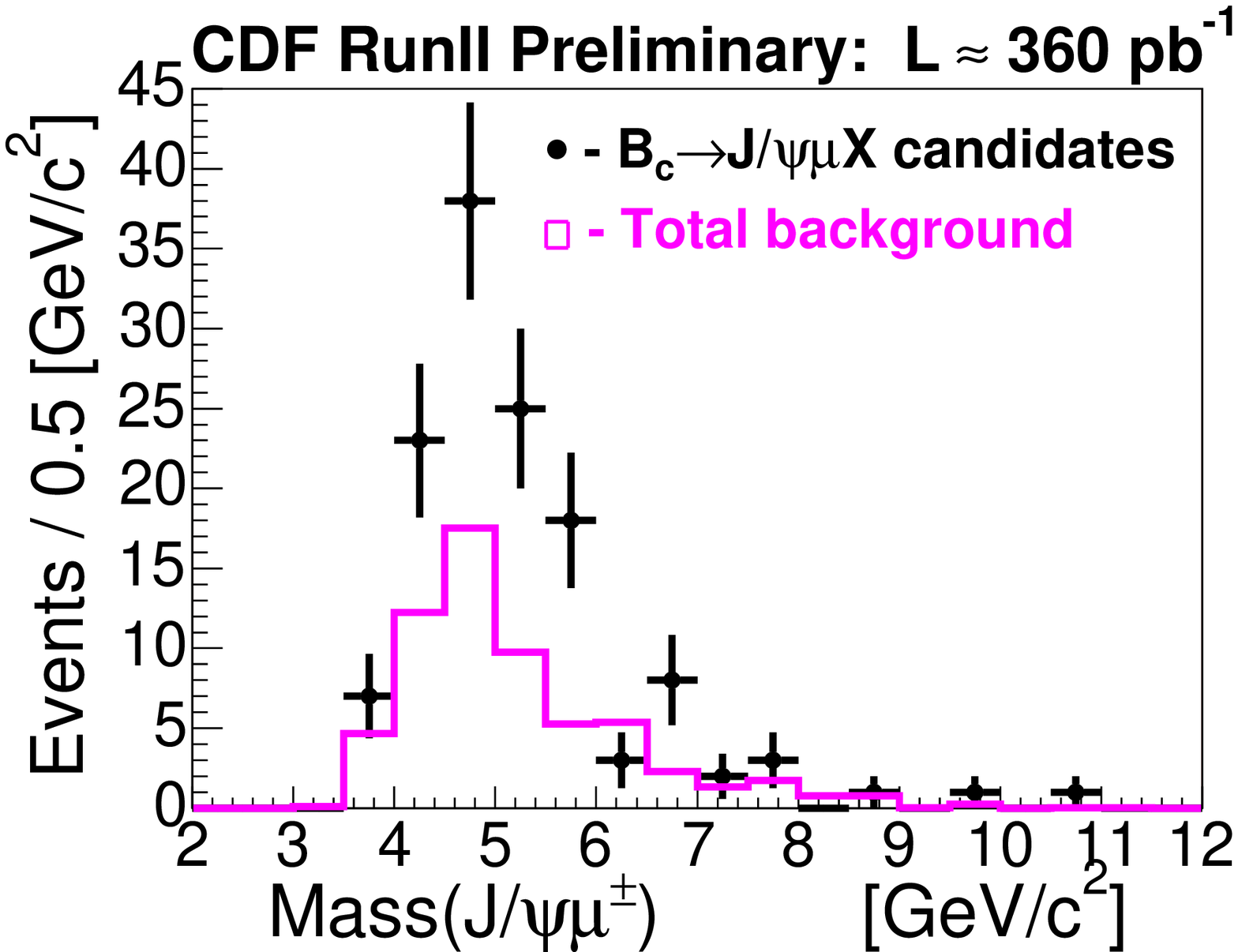}
\efig{5.5}{Signal (crosses are data and red line is Monte Carlo)
and background for the $\bc\rightarrow\jpsi\; e$
  channel}{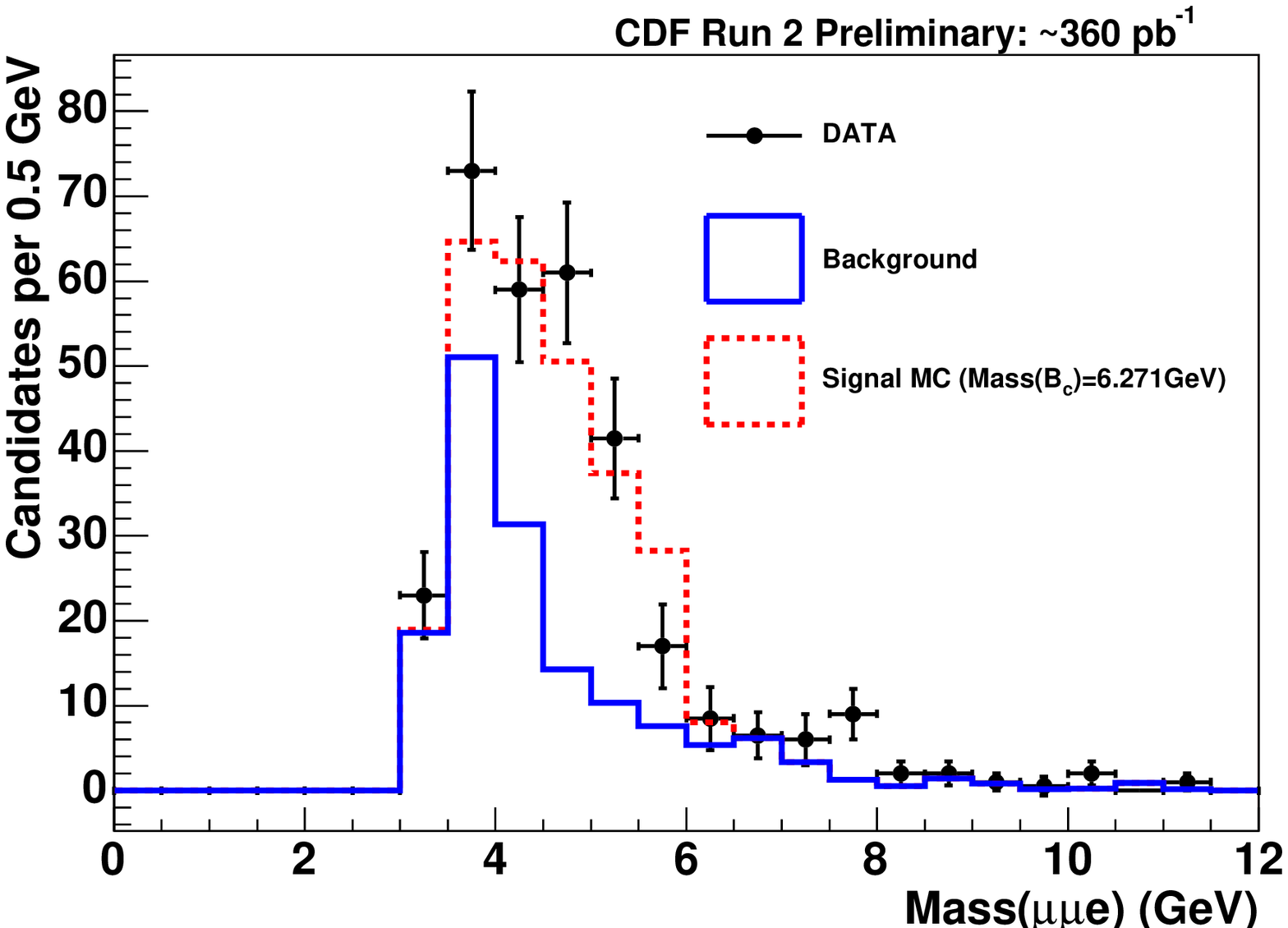}

By carefully evaluating the detection efficiency for signal and
reference channel, we obtain the main result of these measurements,
which is the ratio of branching ratios:
\begin{eqnarray*}
\frac{\sigma(\bc)\times\br{\bc\rightarrow\jpsi\;\mu}}
{\sigma(B)\times\br{B\rightarrow\jpsi\; K}} = \\
0.249 \pm 0.045 {\rm (stat)} ^{+0.107}_{-0.076} {\rm (syst)}
\end{eqnarray*}
and
\begin{eqnarray*}
\frac{\sigma(\bc)\times\br{\bc\rightarrow\jpsi\; e}}
{\sigma(B)\times\br{B\rightarrow\jpsi\; K}} = \\
0.282 \pm 0.038 {\rm (stat)} \pm 0.074 {\rm (syst)}
\end{eqnarray*}

\subsection{Fully reconstructed decay \cite{bc2psipi}}

The $\bc \rightarrow J/\psi \pi$ search was performed using a blind
 analysis method.  The mass values of the $\jpsi \pi$ combinations in
 the search window 5600 $< M(\jpsi \pi) <$ 7200 MeV/$c^2$, referred to
 as $\bc$ candidates, were temporarily hidden.

In order to optimize the significance of a possible signal, we varied
the selection criteria to maximize the function $S_F/(1.5 + \sqrt{B})$
\cite{Punzi}.  Here, $S_F$ is the accepted fraction of signal events,
in this case taken from a Monte Carlo sample, and the background $B$
is the number of accepted $\bc$ candidates.  The following optimized
selection criteria were used: a quality requirement on the $\jpsi \pi$
three-track 3-D vertex fit ($\chi^2 <$ 9 for four degrees of freedom),
a requirement on the pion track contribution to the vertex fit
($\chi_{\pi}^2 <$ 2.6), the impact parameter of the $\bc$ candidate
with respect to the primary vertex ($<$ 65 $\mu$m), the maximum $ct$
where $t$ is the proper decay time of the $\bc$ candidate ($<$ 750
$\mu$m), the transverse momentum of the pion ($>$ 1.8 GeV/$c$), the
3-D angle between the momentum of the $\bc$ candidate and the vector
joining the primary to the secondary vertex ($\beta <$ 0.4 rad), and
the significance of the projected decay length of the $\bc$ candidate
onto its transverse momentum direction ($L_{xy}/\sigma(L_{xy}) >$
4.4). Because of the relatively long $\bc$ lifetime, vertex cuts are
critical in this analysis.

A sample of $B$ mesons, reconstructed in the decay mode $B\rightarrow
\jpsi K$, was analyzed as a control sample in order to check our
understanding of the reconstruction of the relevant variables in the
simulation.

Before ``unblinding'' the $\jpsi \pi$ mass distribution, a procedure
to search for a signal peak was defined. This was based on a scan of
the search region, with a sliding fit window.  We applied the fitting
procedure to the 390 candidates in the unblinded $\jpsi \pi$ mass
distribution.  This provides a value of $\Sigma(m) = S/(1.5 +
\sqrt{B})$ as a function of the mass in the search window.  A maximum
value of $\Sigma_{max} = 3.6$ is found at a mass
$m\approx$6290~MeV/$c^2$

A set of Monte Carlo experiments was performed to determine the
expected distribution of $\Sigma_{max}$ for pure background samples
\cite{Rolke}. The data value of $\Sigma_{max}$ was found to be
exceeded in 0.27\% of Monte Carlo scans

\efig{5.5}{The invariant mass distribution of the $\jpsi \pi$ candidates
and results of an unbinned likelihood fit}{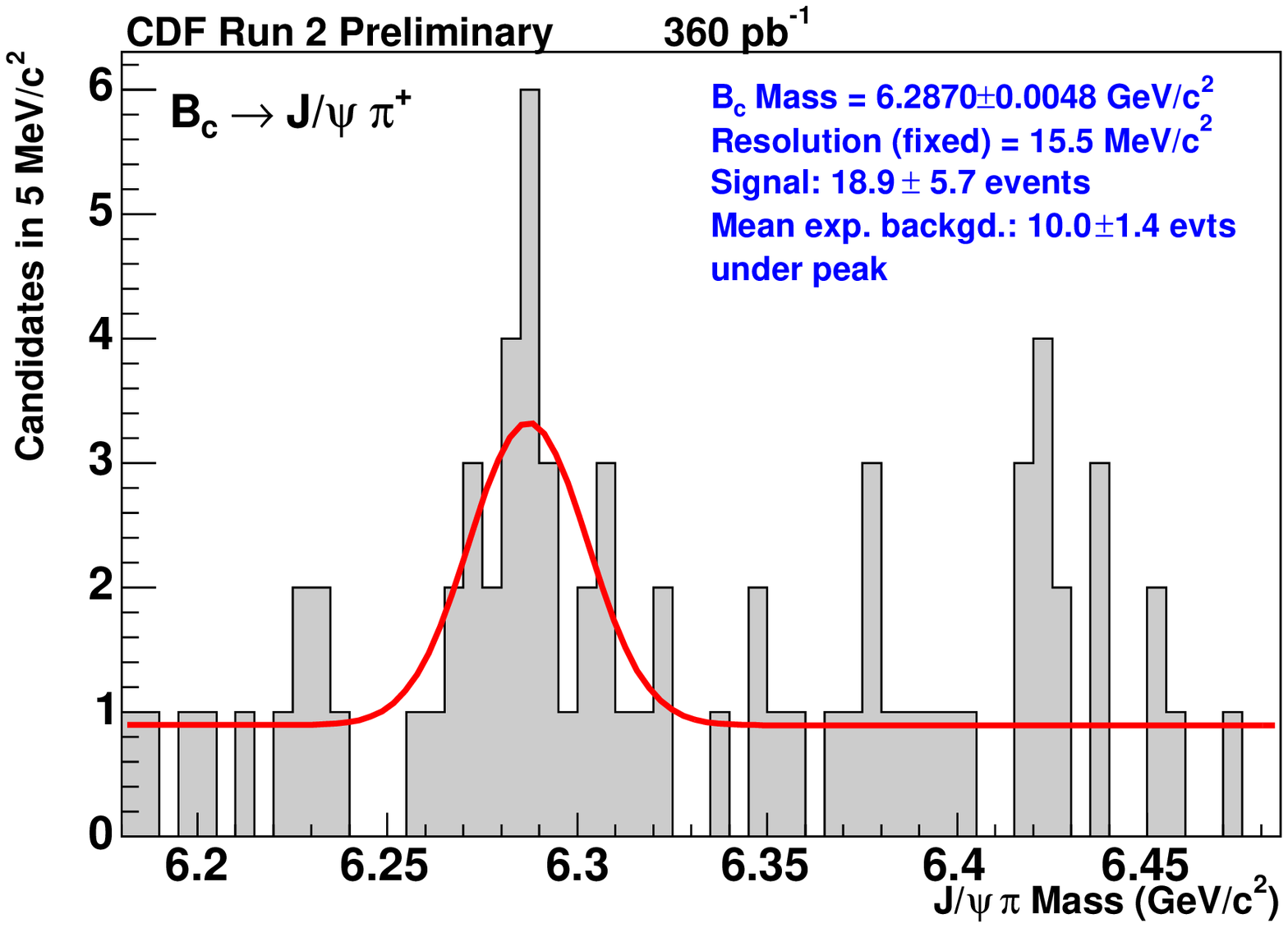}

The exact value of the mass is found with an unbinned likelyhood fit,
shown in figure \ref{fig:Final_Mass_Fit}.  Systematic uncertainties on
the $\bc$ mass determination due to measurement uncertainties on the
track parameters ($\pm 0.3$ MeV/$c^2$), the momentum scale ($\pm 0.6$
MeV/$c^2$), the possible differences in the $p_T$ spectra of the $B$
and $\bc$ mesons ($\pm$0.5 MeV/$c^2$) and our limited knowledge of the
background shape used in the final mass fit ($\pm$0.8 MeV/$c^2$).  The
total systematic uncertainty is evaluated to be $\pm1.1$ MeV/$c^2$.

We observe 18.9 $\pm$ 5.7 signal events on a background of 10.0 $\pm$
1.4 events and the fit to the $\jpsi \pi$ mass spectrum yields a $\bc$
mass of 6287.0 $\pm$ 4.8(stat) $\pm$ 1.1(syst) MeV/$c^2$.

\section{CONCLUSION}

We performed a measurement of mass and ratio of branching fractions of
the $\bc$ meson, on a sample of $360^{-1}$ pb of $p\bar{p}$ collisions
collected at $\sqrt{s}=1.96$ TeV by Collider Detector at Fermilab
(\cdf) at the Tevatron during Run II. Results are compatible with
theory and previous measurements. The Tevatron has recently achieved
1fb$^{-1}$ of data, with which we plan to update
these measurements.

\end{document}